\documentclass[a4paper,reqno,oneside]{article}
\def\Z{\mathbb Z}

\def\pf{\begin{proof}}
\def\pfk{\end{proof}}

\usepackage{amssymb,amsmath,amsthm}

\def\Z{\mathbb Z}

\def\pf{\begin{proof}}
\def\pfk{\end{proof}}

\theoremstyle{theorem}
\newtheorem{lem}{Lemma}[section]

\newtheorem{prop}[lem]{Proposition}
\newtheorem{thm}[lem]{Theorem}

\theoremstyle{definition}
\newtheorem{de}[lem]{Definition}
\newtheorem{pozn}[lem]{Remark}

\begin{document}
\title{ Fine grading of $sl(p^2,\mathbb{C})$ generated by tensor product of
generalized Pauli matrices and its symmetries  }

\author{Edita Pelantov\'a$^*$ \and Milena Svobodov\'a$^*$
\and S\'ebastien Tremblay$^\dag$}

\date{}

\maketitle

\begin{center}
{$^*$Department of Mathematics, FNSPE, Czech Technical University,
Trojanova 13, \\ 120 00 Praha 2, Czech Republic\\
$^\dag$D{\'e}partement de math{\'e}matiques et d'informatique,
Universit{\'e} du Qu{\'e}bec {\`a} Trois-Rivi{\`e}res, \\ C.P. 500
Trois-Rivi\`eres, Qu\'ebec,
G9A 5H7, Canada  \\
$\ $\\
}
\end{center}

\begin{abstract}
Study of the normalizer of the MAD-group corresponding to a fine
grading offers the most important tool for describing symmetries
in the system of non-linear equations connected with contraction
of a Lie algebra. One fine grading that is always present in any
Lie algebra $sl(n,\mathbb{C})$ is the Pauli grading. The MAD-group
corresponding to it is generated by generalized Pauli matrices.
For such MAD-group, we already know its normalizer; its quotient
group is isomorphic to the Lie group $SL(2,\mathbb{Z}_n)\times
\mathbb{Z}_2$.

In this paper, we deal with a more complicated situation, namely
that the fine grading of $sl(p^2, \mathbb{C})$ is given by a
tensor product of the Pauli matrices of the same order $p$, $p$
being a prime. We describe the normalizer of the corresponding
MAD-group and we show that its quotient group is isomorphic to
$Sp(4,\mathbb{F}_p)\times\mathbb{Z}_2$, where $\mathbb{F}_p$ is
the finite field with $p$ elements.
\end{abstract}


\section{Introduction}

\bigskip

A grading $\Gamma$ of a Lie algebra $L$ is a decomposition
$\Gamma: L = \oplus_{i\in J} L_i$  into non-trivial subspaces
$L_i$ such that, for each pair of indices $i,j\in J$, there exists
an index $k\in J$ fulfilling the property $[L_i,L_j]\subset L_k$.
Among all the gradings of a Lie algebra, the most important ones
are fine gradings, since any grading is created from some fine
grading.

It was shown (in \cite{PaZa}) that there is a one-to-one
correspondence between fine gradings of a simple Lie algebra over
$\mathbb{C}$ and maximal Abelian groups of diagonalizable
automorphisms (so called MAD-groups) in ${\cal A}ut \, L$. Each
fine grading of a simple Lie algebra over $\mathbb{C}$ is obtained
as a decomposition of $L$ into eigensubspaces of automorphisms
from a MAD-group. In \cite{HaPaPe}, all MAD-groups of ${\cal A}ut
\, sl(n,\mathbb{C})$ were described.

Let us recall that the Lie algebra $sl(n,\mathbb{C})$ has inner
and outer automorphisms. An inner automorphism is given by a
non-singular matrix $A$ of order $n$ by the prescription $$
Ad_AX:= A^{-1}XA, \ \hbox{for}\ X \in  sl(n,\mathbb{C}).$$ An
outer automorphism is connected with a non-singular matrix $A$ of
order $n$ as well, and it is given by the prescription $$ Out_AX:=
-(A^{-1}XA)^\top, \ \hbox{for}\ X \in  sl(n,\mathbb{C}),$$ where
$A^\top$ denotes the transposed matrix.

An important role in the description of MAD-groups without outer
automorphism is played by generalized Pauli matrices. They were
introduced in \cite{Pauli}:

\medskip
\begin{de}  For a given $n\in \mathbb{N}$, put $\omega=\omega_n =
e^{\frac{2\pi i}{n}}$. A group of matrices
$$ {\cal P}_n := \bigl\{ \omega^jP_n^kQ_n^l\ |\ j,k,l \in \{ 0 ,1,
\ldots, n\!-\!1\}\bigr\}, \  \hbox{where}\ $$
$$ P_n = \left( \begin{array}{llllc}
1&0&0&\ldots& 0\\
0&\omega&0&\ldots & 0\\
0&0&\omega2&\ldots &0 \\
\vdots & & & \ddots& \\
0& 0 & 0&  \ldots  &\omega ^{n-1}
\end{array}\!\!\!
\right)\in \mathbb{C}^{n\times n}\quad {\rm and }\quad Q_n =
\left( \begin{array}{llllc}
0&1&0&\ldots& 0\\
0&0&1&\ldots & 0\\
\vdots & & &  \ddots& \\
0&0&0&\ldots &1 \\
1& 0 & 0& \ldots  &0
\end{array}\!\!\!
\right)\in \mathbb{C}^{n\times n},  $$ is called the {\bf Pauli
group} of order $n$; $P_n$ and $Q_n$ are the generalized Pauli
matrices of order $n$.
\end{de}

\noindent Let us note that the matrices $P_n$ and $Q_n$ do not
commute, since $Q_nP_n=\omega P_nQ_n$. Nevertheless, the inner
automorphisms  corresponding to these matrices do commute: $
Ad_{Q_n} Ad_{P_n} =  Ad_{P_n}Ad_{Q_n}$.

\medskip

 \noindent In order to describe  MAD-groups of the algebra
$sl(n, \mathbb{C})$, we need further notation:

\noindent  The group of non-singular diagonal matrices of order
$n$ will be
 denoted by ${\cal D}_n$, i.e.
 $${\cal D}_n = \bigl\{\mathrm{diag}(\alpha_1, \alpha_2, \ldots ,\alpha_n)\ |\
 \alpha_1, \alpha_2, \ldots, \alpha_n \in  \mathbb{C}\setminus\{0\}\bigr\}.$$

 \noindent If  $G_1$ and $G_2$  are groups of matrices, then  $
 G_1\otimes G_2$ denotes the group of all tensor
 products\footnote{If $A\in\mathbb{C}^{n\times n}$ and  $B \in \mathbb{C}^{m\times m}$, then
 the tensor product $A\otimes B \in \mathbb{C}^{nm\times nm}$  is
defined by $(A\otimes B)_{IJ}= A_{i_1i_2}B_{j_1j_2}$, where
$i_1,i_2 \in \{0,1,\ldots, n-1\}$, $j_1,j_2 \in \{0,1,\ldots ,
m-1\}$, $I,J\in \{0,1,\ldots ,mn-1\}$ and $I=i_1m+j_1$,
$J=i_2m+j_2$.}
 $A\otimes B$, where $A\in G_1$ and $B\in G_2$.

\medskip

The MAD-groups of ${\cal A}ut \, sl(n,\mathbb{C})$ can be divided
into two classes, depending whether or not they contain an outer
automorphism. It is proved in \cite{HaPaPe} that any MAD-group in
the automorphism group ${\cal A}ut\, sl(n, \mathbb{C})$ containing
only inner automorphisms is isomorphic to a group of the following
form $$ {\cal G} = \bigl\{ Ad_A\ |\ A\in {\cal P}_{n_1} \otimes
{\cal P}_{n_2} \otimes \ldots \otimes {\cal P}_{n_{r-1}} \otimes
{\cal D}_{n_r}\bigr\},$$ where $n_1n_2\ldots n_r=n$ and $n_{i-1}$
divides $n_i$ for any $i=2,3,\ldots,  r-1$.

\bigskip

A grading $\Gamma: L = \oplus_{i\in J} L_i$ of a Lie algebra $L$
is a starting point for searching for graded contractions of the
Lie algebra. This method for finding  contractions of Lie algebras
was used by several authors \cite{Win, MoPa, MuPa}. In this type
of contraction, we define new Lie brackets by prescription $$ [
x,y]_{new} := \varepsilon_{jk}[x,y], \ \hbox{where}\ x\in L_j,
y\in L_k.$$ The complex or real parameters $\varepsilon_{jk}$, for
$j,k\in J$, must be determined in such way that the vector space
$L$ with the binary operation $[.,.]_{new}$ forms again a Lie
algebra. Antisymmetry of Lie brackets  demands that
$\varepsilon_{jk} = \varepsilon_{kj}$. Compliance with the Jacobi
identity, however, already implies that the coefficients
$\varepsilon_{jk}$  fulfill a complicated system of quadratic
equations, which is in general difficult to solve. For description
of symmetries of this system, it is important to know the
symmetries of the original grading $\Gamma$. By a symmetry of a
grading  of the Lie algebra $L$ we mean such an automorphism $g\in
{\cal A}ut\, L$ that
\begin{equation}\label{symetrie} \hbox{for each}\ j\in J\
\hbox{there exists}\ k\in J\ \hbox{fulfilling}\  gL_j=L_k.
\end{equation}
Let us suppose that a fine grading $\Gamma: sl(n, \mathbb{C}) =
\oplus_{j\in J}L_j$ corresponds to a MAD-group ${\cal G}  \subset
{\cal A}ut\, sl(n,\mathbb{C})$. It means that
\begin{equation}\label{MAD}
hL_k=L_k\  \ \hbox{for all}\ h\in {\cal G} \ \hbox{and}\ \ k\in J.
\end{equation}
Combining (\ref{MAD}) and (\ref{symetrie}), we obtain $$ ghg^{-1}
L_j = L_j \ \ \hbox{for any} \ \ j\in J.$$ The maximality of
${\cal G}$ implies that $ghg^{-1}\in {\cal G}$ for any $h\in {\cal
G}$. This means, in other words, that the symmetries of the
grading $\Gamma$ corresponding to the MAD-group ${\cal G}$ form a
group $$ {\cal N}({\cal G}) = \{ g\in {\cal A}ut\,
sl(n,\mathbb{C})\ |\ g{\cal G}g^{-1} \subseteq \cal G\}.$$
 This group is usually called the {\bf normalizer}\ of the subgroup ${\cal
 G}$ in ${\cal A}ut\,
sl(n,\mathbb{C})$.\\

 The definition of the normalizer ${\cal N}({\cal G})$ implies
 that ${\cal G} \subset {\cal N}({\cal G})$. Moreover, $\cal G$ is
 a normal subgroup of ${\cal N}({\cal G})$. Thus, when describing
 ${\cal N}({\cal G})$, it is sufficient to inspect the quotient
 group ${\cal N}({\cal G}){/\cal G}$.\\

 The article \cite{HaPaPeTo} studied the normalizer of one
 MAD-group of $sl(n,\mathbb{C})$, namely $${\cal G} = \{Ad_A\ |\ A\in
 {\cal P}_n\}.$$
It was shown that ${\cal N}({\cal G}){/\cal G}$ is isomorphic to
the matrix group $\{ A\in \mathbb{Z}_n^{2\times 2}\  |\ \det A
=\pm 1\}$, $\mathbb{Z}_n$ being a cyclic group of order $n$. This
result was used for obtaining all the graded contractions (see
\cite{HrNoTo}) of Lie algebra $sl(3,\mathbb{C})$ that arise from
the Pauli grading
 $$ sl(3,\mathbb{C}) = \oplus_{(j,k)\in J} \{P_3^jQ_3^k\}_{lin}, \
 \ J=\mathbb{Z}_3\times \mathbb{Z}_3 \setminus \{(0,0)\}.$$

\medskip

In this article, we are going to study the symmetries  of the
grading corresponding to the MAD-group
\begin{equation}\label{odted}
{\cal G} = \{Ad_A\ |\ A\in
 {\cal P}_n\otimes {\cal P}_n \}\subset {\cal A}ut\, sl(n^2, \mathbb{C}), \
 n \ \hbox{prime}.
 \end{equation}
In the sequel we will use notation $P$ and $Q$ instead of $P_n$
and $Q_n$, and by the letter ${\cal G}$
 we will denote only the group given by (\ref{odted}).

 \medskip

If $n$ and $m$  are coprime integers, then the tensor product
${\cal P}_n\otimes {\cal P}_m$ of the Pauli groups  ${\cal P}_n$
and ${\cal P}_m$ is isomorphic to the Pauli group
 ${\cal P}_{nm}$, therefore  it is a natural step in investigation
 of symmetries of gradings to devote attention to the MAD-group
 given by $ {\cal P}_n\otimes {\cal P}_n $.

\section{The normalizer of the MAD-group corresponding to the
tensor product  ${\cal P}_n \otimes   {\cal P}_n $  }

If $(g_i)_{i\in I}$ is a set of generators of a group $ {\cal H}
\subset {\cal A}ut\, L$, then $\varphi \in {\cal A}ut\, L$ belongs
to the normalizer ${\cal N}({\cal H})$ if and only if $\varphi
g_i\varphi^{-1} \in {\cal  H}$ for all the generators $g_i$. Since
 $ {\cal P}_n\otimes {\cal P}_n  = \{ P^iQ^j\otimes P^kQ^l\}$,
our MAD-group $${\cal G} = \{ Ad_{P^iQ^j\otimes P^kQ^l}\ |\
i,j,k,l\in \mathbb{Z}_n\}$$ has four generators (we use the letter
$I$ for unit matrix of order $n$):

\medskip

\centerline{$Ad_{A_1}$, where $A_1 = P\otimes I$,}

\centerline{ $Ad_{A_2}$, where $A_2 =Q\otimes I$,}
 \centerline{$Ad_{A_3}$, where $A_3 = I\otimes P$,}

 \centerline{ $Ad_{A_4}$, where $A_4= I\otimes Q$.}

 \medskip

\noindent Any element  of the MAD-group ${\cal G}$ is
characterized by a quadruple of indices in $\mathbb{Z}_n$. We know
that an automorphism $\varphi \in {\cal A}ut \,sl(n^2,\mathbb{C})$
belongs to ${\cal N}({\cal G})$ if and only if $\varphi Ad_{A_i}
\varphi^{-1}  \in {\cal G}$ for $i=1,2,3,4$. Thus each $\varphi
\in {\cal N}({\cal G})$ is characterized by a set of 16
coefficients $(a_{ij})_{i,j=1}^4$ such that $$\varphi Ad_{A_j}
\varphi^{-1} = Ad_{P^{a_{1j}}Q^{a_{2j}}\otimes
P^{a_{3j}}Q^{a_{4j}}}=
Ad_{A_1^{a_{1j}}A_2^{a_{2j}}A_3^{a_{3j}}A_4^{a_{4j}}}\quad
\hbox{for }\  j = 1,2,3,4.$$
 We order these 16 parameters into a matrix $C(\varphi) \in
 \mathbb{Z}_n^{4\times 4}$ as follows:
\begin{equation}\label{mapping}
\varphi \mapsto C(\varphi) = \left( \begin{array}{llll} a_{11}&
a_{12}&
 a_{13} & a_{14}\\
a_{21}& a_{22}&
 a_{23} & a_{24}\\
 a_{31}& a_{32}&
 a_{33} & a_{34}\\
 a_{41}& a_{42}&
 a_{43} & a_{44}
  \end{array}
 \right).
 \end{equation}

\begin{pozn}\label{coseslije} Obviously, the assignment  $\varphi \mapsto C(\varphi)$ implies
that $C(\varphi) = I_4$ if and only if $\varphi$ commutes with
each generator of the MAD-group ${\cal G}$, and thus with the
whole MAD-group. This means that $\varphi$ necessarily belongs to
${\cal G}$ (due to the maximality of ${\cal G}$). Shortly, we have
\begin{equation}\label{jednicka}
C(\varphi)=I_4 \quad \Longleftrightarrow \quad \varphi \in {\cal
G}.
\end{equation}
\end{pozn}
The advantage of such ordering of the 16 coefficients
corresponding to $\varphi \in {\cal N}({\cal G})$ is obvious from
the following statement:

\begin{prop}\label{soucin} Let $\varphi, \psi \in {\cal N}({\cal G})$. Then  $C(\varphi\psi)
= C(\varphi)C(\psi).$
\end{prop}

\pf We denote the coefficient  matrices by $C(\varphi) =
(a_{ij})_{i,j=1}^4$, and $C(\psi)= (b_{ij})_{i,j=1}^4$. Let us
apply the automorphism $\varphi\psi$ on an element $Ad_{A_p} \in
{\cal G}$ as follows: $$
\begin{array}{lll}
&&(\varphi\psi )Ad_{A_p}(\varphi\psi)^{-1}= \varphi(\psi
Ad_{A_p}\psi^{-1})\varphi^{-1} = \varphi(
Ad_{A_1^{b_{1p}}A_2^{b_{2p}}A_3^{b_{3p}}A_4^{b_{4p}}})\varphi^{-1}\\*[2ex]
&=& (\varphi Ad_{A_4}\varphi^{-1})^{b_{4p}}(\varphi
Ad_{A_3}\varphi^{-1})^{b_{3p}} (\varphi
Ad_{A_2}\varphi^{-1})^{b_{2p}}(\varphi
Ad_{A_1}\varphi^{-1})^{b_{1p}}
\\*[2ex] &=&
(Ad_{A_1^{a_{14}}A_2^{a_{24}}A_3^{a_{34}}A_4^{a_{44}}})^{b_{4p}}
(Ad_{A_1^{a_{13}}A_2^{a_{23}}A_3^{a_{33}}A_4^{a_{43}}})^{b_{3p}}
(Ad_{A_1^{a_{12}}A_2^{a_{22}}A_3^{a_{32}}A_4^{a_{42}}})^{b_{2p}}
(Ad_{A_1^{a_{11}}A_2^{a_{21}}A_3^{a_{31}}A_4^{a_{41}}})^{b_{1p}}.
\end{array}
$$ Since $A_jA_k = const\, A_kA_j$, we have $Ad_{A_jA_k}
=Ad_{A_kA_j}$for any $j,k=1,2,3,4.$ Therefore $$ (\varphi\psi
)Ad_{A_p}(\varphi\psi)^{-1} =
Ad_{A_1^{c_{1p}}A_2^{c_{2p}}A_3^{c_{3p}}A_4^{c_{4p}}},$$ where
\begin{eqnarray*}
c_{1p} &= &a_{11}b_{1p}+a_{12}b_{2p}+a_{13}b_{3p}+a_{14}b_{4p}\\
c_{2p}&= &a_{21}b_{1p}+a_{22}b_{2p}+a_{23}b_{3p}+a_{24}b_{4p}\\
c_{3p}&=&a_{31}b_{1p}+a_{32}b_{2p}+a_{33}b_{3p}+a_{34}b_{4p}\\
c_{4p}&=&a_{41}b_{1p}+a_{42}b_{2p}+a_{43}b_{3p}+a_{44}b_{4p}.
\end{eqnarray*}
This means, in brief notation, that $C(\varphi\psi) =
C(\varphi)C(\psi)$. \pfk

\bigskip

We prove below that the matrix $C(\varphi)$ assigned to the
element $\varphi$ of the normalizer ${\cal N}({\cal G})$
characterizes a coset belonging  to the quotient group ${\cal
N}({\cal G})/{\cal G}$:

\medskip

\begin{prop}\label{prop factor G} Let $\varphi, \psi$ belong  to the normalizer ${\cal N}({\cal
G})$ of the MAD-group ${\cal G}$. Then $C(\varphi) = C(\psi)$ if
and only if there exists $h \in {\cal G}$ such that $\varphi =
h\psi$.
\end{prop}

\pf Let $\varphi, \psi\in {\cal N}({\cal G})$ such that
$C(\varphi)=C(\psi)$. Since  $\psi^{-1} \in {\cal N}({\cal G})$ as
well, we obtain from Proposition \ref{soucin} $$ C(\varphi
\psi^{-1}) = C(\varphi)C(\psi^{-1}) = C(\psi)C(\psi^{-1})=
C(\psi\psi^{-1}) = C(Id) = I_4. $$ By Remark \ref{coseslije},
$\varphi\psi^{-1}$ commutes with all elements of ${\cal G}$, which
is only possible when $\varphi\psi^{-1} \in {\cal G}$.

The opposite implication  follows directly from Remark
\ref{coseslije}. \pfk

\begin{lem} The outer automorphism $Out_I$ belongs to the normalizer
${\cal N}({\cal G})$,
 and
 $$C(Out_I) = \mathrm{diag}(-1,1,-1,1).$$
\end{lem}

\pf Let us denote $\varphi_0=Out_I$. As $\varphi_0$ is given by
the prescription $\varphi_0X = -X^\top$, clearly $\varphi_0^{-1} =
\varphi_0$. We can derive for any inner automorphism $Ad_A$ that
$$(\varphi_0Ad_A\varphi_0^{-1})(X) = (\varphi_0 Ad_A)(-X^\top) =
\varphi_0(-A^{-1}X^\top A) $$
 $$= (A^{-1}X^\top A)^\top = A^\top X
A^{-\top} = (A^{-\top})^{-1} X A^{-\top} = Ad_{A^{-\top}}(X),$$
where we used abbreviated  notation $A^{-\top}$ instead of
$(A^{-1})^\top$. This notation is used in the sequel as well.

 Thus we have shown that the action of $Out_I$ on any
inner automorphism $Ad_A$ is
 \begin{equation}\label{vnejsi}
 Out_I Ad_A Out_I^{-1} = Ad_{A^{-\top}}.
 \end{equation}
Now, for each generator $Ad_{A_j}$, $j=1,2,3,4$, of the MAD-group
${\cal G}$, we prove that $\varphi_0Ad_{A_j}\varphi_0^{-1}$
belongs to ${\cal G}$:

Let us recall the following trivial properties of matrices $P$ and
$Q$ and the properties of tensor product:

~~i) \ $(A\otimes B)^{-1} = A^{-1}\otimes B^{-1}$;

~ii) \ $(A\otimes B)^{\top} = A^\top\otimes B^\top$;

iii) \ $P^\top = P$ , as $P$ is diagonal;

iv) \ $Q^{-1} = Q^\top$, as $Q$ is a permutation matrix.

\medskip

\noindent Using these relations, we obtain:

1. \ $ A_1^{-\top}= (P\otimes I)^{-\top} = P^{-\top}\otimes I =
P^{-1} \otimes I = (P\otimes I)^{-1} = A_1^{-1}$;

2. \ $ A_2^{-\top}= (Q\otimes I)^{-\top} = Q^{-\top}\otimes I =
Q\otimes I = A_2$;

3. \ $ A_3^{-\top}= (I\otimes P)^{-\top} = I\otimes P^{-\top} =
I\otimes P^{-1} = (I\otimes P)^{-1} = A_3^{-1}$;

4. \ $ A_4^{-\top}= (I\otimes Q)^{-\top} = I\otimes Q^{-\top} =
I\otimes Q  = A_4$.

\noindent  Statements 1.-4. together with equation (\ref{vnejsi})
already prove the lemma. \pfk

\begin{pozn}\label{rozklad}  Product of two outer automorphisms is an inner automorphism.
Thus, when describing the set of all automorphisms in ${\cal
N}({\cal G})$, we can focus on the subgroup ${\cal N}_{in}({\cal
G})$ containing all the inner automorphisms in ${\cal N}({\cal
G})$. The whole ${\cal N}({\cal G})$ can then be described as
 $$ {\cal N}({\cal G}) = {\cal N}_{in}({\cal G}) \cup Out_I\, {\cal N}_{in}({\cal
 G}).$$
\end{pozn}

The following theorem shows the connection between normalizers of
these MAD-groups and the symplectic groups over finite field which
were introduced in \cite{Artin}.
\begin{thm}  Let $Ad_A$ be an inner
 automorphism contained in the normalizer ${\cal N}({\cal G})$ of the MAD-group
 ${\cal G}$, and let $C(Ad_A)$ be the coefficient matrix corresponding to $Ad_A$.  Then
\begin{equation}\label{symplectic} C(Ad_A)  \in Sp\,(4, \mathbb{Z}_n) := \{ X\in
\mathbb{Z}_n^{4\times 4}\ |\ X^\top\! JX =
 J\},
 \end{equation}
where $J= \left(\begin{smallmatrix}
      0&1&0&0\\-1&0&0&0\\
      0&0&0&1\\
      0&0&-1&0\end{smallmatrix}\right)  = I_2\otimes \left(\begin{smallmatrix}
      0&1\\-1&0\end{smallmatrix}\right).$
\end{thm}

\pf  Let us denote $C(Ad_A) = (a_{ij})_{i,j=1}^4$. The definition
of the matrix   $C(Ad_A)$ implies that
\begin{equation}\label{(i)}
 Ad_AAd_{A_p}(Ad_A)^{-1} = Ad_{A^{-1}A_pA} =
Ad_{A_1^{a_{1p}}A_2^{a_{2p}}A_3^{a_{3p}}A_4^{a_{4p}}} \quad
\hbox{for } \ p=1,2,3,4.
\end{equation}
As $Ad_K = Ad_H$ if and only if $K=\alpha H$ for some $\alpha \in
\mathbb{C}-\{0\}$, we obtain from (\ref{(i)}) existence of four
non-zero constants $\alpha_p, p=1,2,3,4$, such that
\begin{equation}\label{(ii)}
A_p=\alpha_pA A_1^{a_{1p}}A_2^{a_{2p}}A_3^{a_{3p}}A_4^{a_{4p}}
A^{-1}.
\end{equation}

\noindent We derive easily from the basic relation $QP=\omega PQ$
that
\begin{eqnarray}\label{(1)}
             A_1A_2&=&\omega^{-1}A_2A_1\\
            \label{(2)} A_3A_4&=&\omega^{-1}A_4A_3.
\end{eqnarray}
The remaining pairs $A_i,A_j$ commute:
\begin{eqnarray}\label{(3)}
             A_1A_3&=&A_3A_1\\
\label{(4)}  A_1A_4&=&A_4A_1\\ \label{(5)}  A_2A_3&=&A_3A_2\\
\label{(6)}  A_2A_4&=&A_4A_2.
\end{eqnarray}

\noindent By inputting $A_1, A_2$ expressed  in  the form
(\ref{(ii)}) into the relation (\ref{(1)}), we obtain
$$
\begin{array}{ll}
& \alpha_1AA_1^{a_{11}}A_2^{a_{21}}A_3^{a_{31}}A_4^{a_{41}}
A^{-1}\alpha_2AA_1^{a_{12}}A_2^{a_{22}}A_3^{a_{32}}A_4^{a_{42}}
A^{-1} \\*[2ex] =&
\omega^{-1}\alpha_2AA_1^{a_{12}}A_2^{a_{22}}A_3^{a_{32}}A_4^{a_{42}}
A^{-1}\alpha_1AA_1^{a_{11}}A_2^{a_{21}}A_3^{a_{31}}A_4^{a_{41}}
A^{-1},
\end{array}
$$
and, after simplification and using relations
(\ref{(1)})-(\ref{(6)}),
$$
\begin{array}{ll}
&\omega^{a_{21}a_{12}+a_{41}a_{32}}A_1^{a_{11}+a_{12}}A_2^{a_{21}+a_{22}}
A_3^{a_{31}+a_{32}}A_4^{a_{41}+a_{42}}\\*[2ex] =& \omega^{-1
+a_{22}a_{11} +a_{42}a_{31}}A_1^{a_{11}+a_{12}}A_2^{a_{21}+a_{22}}
A_3^{a_{31}+a_{32}}A_4^{a_{41}+a_{42}}.
\end{array}
$$
This implies that $\omega^{a_{21}a_{12}+a_{41}a_{32}} =\omega^{-1
+a_{22}a_{11} +a_{42}a_{31}}$, and therefore
 \begin{eqnarray}\label{(I)}
  1&=&a_{11}a_{22}- a_{21}a_{12} + a_{31}a_{42}- a_{41}a_{32}\ \
  ({\rm{mod}}\ n).
  \end{eqnarray}
Analogously, the equations (\ref{(2)})--(\ref{(6)}) result in
\begin{eqnarray}\label{(II)}
  1&=&a_{13}a_{24}- a_{14}a_{23} + a_{33}a_{44}- a_{34}a_{43}\ \
  ({\rm{mod}}\ n)\\
  \label{(III)}
  0&=&a_{11}a_{23}- a_{13}a_{21} + a_{31}a_{43}- a_{33}a_{41}\ \
  ({\rm{mod}}\ n)\\
  \label{(IV)}
  0&=&a_{11}a_{24}- a_{14}a_{21} + a_{31}a_{44}- a_{34}a_{41}\ \
  ({\rm{mod}}\ n)\\
\label{(V)}
  0&=&a_{12}a_{23}- a_{13}a_{22} + a_{32}a_{43}- a_{33}a_{42}\ \
  ({\rm{mod}}\ n)\\
\label{(VI)}
  0&=&a_{12}a_{24}- a_{14}a_{22} + a_{32}a_{44}- a_{34}a_{42}\ \
  ({\rm{mod}}\ n).
  \end{eqnarray}

It can be easily verified (by a direct  calculation) that the
matrix $(a_{ij})_{i,j=1}^4$ belongs to the group
$Sp(4,\mathbb{Z}_n)$ if and only if the matrix elements $a_{ij}$
fulfill equations (\ref{(I)})--(\ref{(VI)}).\pfk

\begin{pozn} We were notified \cite{ustne} that the set $Sp\,(4,
\mathbb{Z}_n)$  defined analogously to (\ref{symplectic}) is a
group even in the case  when $\mathbb{Z}_n$ is not a field. All
our previous considerations hold therefore for any positive
integer $n$. But our deductions in the sequel already need $n$ to
be a prime number.
\end{pozn}

We are going to prove that the mapping given by (\ref{mapping}) is
in fact  the mapping on the whole symplectic group $Sp(4,\mathbb
F_n)$. To show it we need to find for any element of $Sp(4,\mathbb
F_n)$ its preimage, or equivalently for any generator of
$Sp(4,\mathbb F_n)$ its preimage.

To simplify the proof we need to find the smallest possible set of
generators of $Sp(4,\mathbb F_n)$.  In \cite{Markus} a set of
generators of the group $Sp(2m, K)$ over a finite field $K$ is
described. In case $m=2$, the set of generators contains
$n^4-n^2+n+1$ elements, where $n$ is the cardinality of $K$. As we
show in Appendix B, it is possible to reduce the number of
generators of the group $Sp(4,\mathbb F_n)$   to four matrices. In
formal notation{\footnote{We recall that for elements $k_1, k_2,
\ldots, k_s$ of a group $G$, the notation $\langle k_1,k_2,
\ldots, k_s\rangle$ means the smallest subgroup of the group $G$
containing $k_1, k_2, \ldots, k_s$.}},
$$Sp(4,\mathbb F_n)=\langle D_1, D_2, D_3, D_4\rangle,$$
where
\begin{equation}\label{maticeD}
D_1 = \left( \begin{smallmatrix}
 1& 1& 0& 0\\
 0& 1& 0& 0\\
 0& 0& 1& 0\\
 0& 0& 0& 1\\
  \end{smallmatrix}
 \right), \ \
D_2 = \left( \begin{smallmatrix}
 0& 1& 0& 0\\
 -1& 0& 0& 0\\
 0& 0& 1& 0\\
 0& 0& 0& 1\\
  \end{smallmatrix}
 \right), \ \
 D_3 = \left( \begin{smallmatrix}
 0& 0& 1& 0\\
 0& 0& 0& 1\\
 1& 0& 0& 0\\
 0& 1& 0& 0\\
  \end{smallmatrix}
 \right), \ \  {\rm and} \ \
 D_4 = \left( \begin{smallmatrix}
 1& 0& 0& 0\\
 0& 1& 0& -1\\
 1& 0& 1& 0\\
 0& 0& 0& 1\\
  \end{smallmatrix}
 \right).
 \end{equation}
We show that these four matrices are images of inner automorphisms
which belong to the normalizer of the group $\mathcal G$.


\begin{prop}\label{theo_inverse images}
Let $n$ be a prime. Then for the four matrices $D_j$ as introduced
in (\ref{maticeD}), there exist inner automorphisms
$\varphi_j=Ad_{B_j}\in\mathcal N _{in}(\mathcal G)$ such that
$D_j=C(\varphi_j)=C(Ad_{B_j})$.
\end{prop}


 The proof is postponed to the Appendix A since it is rather
technical and we do not want to interrupt coherency in the content
of the article.

\medskip

\noindent The immediate consequence of the previous proposition is
the following main result of the article.

\bigskip

\begin{thm}
Let $n$ be a prime. The mapping $\varphi \mapsto C(\varphi)$
defined in (\ref{mapping}) is an isomorphism between groups $$
{\cal N}_{in}({\cal G})/{\cal G}\simeq Sp(4,\mathbb{F}_n)=\{ X\in
\mathbb{Z}_n^{4\times 4}\ |\ X^\top\! JX =
 J\},$$

 $$ and  \qquad  {\cal
N}({\cal G})/{\cal G}\simeq \{ X\in \mathbb{Z}_n^{4\times 4}\ |\
X^\top\! JX =
 \pm J\},$$

\noindent where $J= \left(\begin{smallmatrix}
      0&1&0&0\\-1&0&0&0\\
      0&0&0&1\\
      0&0&-1&0\end{smallmatrix}\right)  = I_2\otimes \left(\begin{smallmatrix}
      0&1\\-1&0\end{smallmatrix}\right).$
\end{thm}
\pf
\begin{itemize}
\item The mapping $\varphi \mapsto C(\varphi)$ from ${\cal
N}_{in}({\cal G})/{\cal G}$ to $Sp(4,\mathbb{F}_n)$ is a
homomorphism, as $C(\varphi\psi)=C(\varphi)C(\psi)$, which was
proved in Proposition~\ref{soucin}.
 \item The mapping $\varphi
\mapsto C(\varphi)$ from ${\cal N}_{in}({\cal G})/{\cal G}$ to
$Sp(4,\mathbb{F}_n)$ is injective, as shown in
Proposition~\ref{prop factor G}. \item The group
$Sp(4,\mathbb{F}_n)$ is generated by four matrices
$D_1,D_2,D_3,D_4$ (see Theorem~\ref{theo_D1234} in Appendix).
\item All the matrices $D_j$, $j=1,2,3,4$, have their inverse
images $\varphi_j\in {\cal N}_{in}({\cal G})/{\cal G}$, such that
$C(\varphi_j)=D_j$ (see Proposition~\ref{theo_inverse images}).
This implies that the mapping $\varphi \mapsto C(\varphi)$ is also
surjective.
\end{itemize}
In total, we see that the mapping $\varphi \mapsto C(\varphi)$ is
an isomorphism from ${\cal N}_{in}({\cal G})/{\cal G}$ onto
$Sp(4,\mathbb{F}_n)$.

\medskip

To show isomorphism between ${\cal N}({\cal G})/{\cal G}$ and
$Sp(4,\mathbb{F}_n) \otimes \mathbb{Z}_2$, it is enough to use
Remark \ref{rozklad} and the fact that  the matrix $ M:= C(Out_I)
= diag(-1,1,-1,1)$ corresponding to the outer automorphisms $Out_I
\in {\cal N}({\cal G})$ satisfies the equality  $M^\top J M = -J$.

 \pfk


\section{Conclusions}

Let us summarize the content of the article:

\begin{enumerate}
\item The normalizer $\mathcal N(\mathcal G)$ of the MAD-group
$\mathcal G=\{Ad_{P^iQ^j\otimes P^kQ^l}\,|\, i,j,k,l\in\mathbb
Z_n\}\subset {\cal A}ut\,sl(n^2,\mathbb C)$ consists of two
subsets:
\begin{eqnarray*}
\mathcal N_{in}(\mathcal G) & \quad & \hbox{the group of all inner
automorphisms in}\ \mathcal N(\mathcal G),\\ Out_I\mathcal
N_{in}(\mathcal G) & \quad & \hbox{the set of all outer
automorphisms in}\ \mathcal N(\mathcal G).
\end{eqnarray*}

\item We provide an explicit expression of the four generators of
$\mathcal N_{in}(\mathcal G)/\mathcal G$, namely the inner
automorphisms $Ad_{B_j}, j=1,2,3,4$ (see
Proposition~\ref{theo_inverse images}).

\item Altogether, we can write the set of generators
\begin{eqnarray*}
\mathcal G&=&\langle Ad_{P\otimes I}, Ad_{Q\otimes I},
Ad_{I\otimes P}, Ad_{I\otimes Q}\rangle,\\
\mathcal N_{in}(\mathcal G)/\mathcal G&=&\langle Ad_{B_1},
Ad_{B_2}, Ad_{B_3}, Ad_{B_4}\rangle,\\
\mathcal N(\mathcal G)&=&\mathcal N_{in}(\mathcal G)\cup
Out_I\mathcal N_{in}(\mathcal G).
\end{eqnarray*}

\item Thus, one can generate each element of the normalizer from
the set $\{Ad_{B_1},$ $Ad_{B_2},$ $Ad_{B_3},$ $Ad_{B_4},$ $Out_I,$
$Ad_{P\otimes I},$ $Ad_{Q\otimes I},$ $Ad_{I\otimes P},$
$Ad_{I\otimes Q}\}$. In formal notation, $$ \mathcal N(\mathcal
G)=\langle Ad_{B_1}, Ad_{B_2}, Ad_{B_3}, Ad_{B_4}, Out_I,
Ad_{P\otimes I}, Ad_{Q\otimes I}, Ad_{I\otimes P}, Ad_{I\otimes
Q}\rangle,$$ where the matrices $B_j$ were defined in the proof of
Proposition~(\ref{theo_inverse images}).

\end{enumerate}

The description of the normalizer as done in this article was only
possible for $n$ prime. For $n$ non-prime, the problem is still
open.

It was shown previously that for the Pauli grading, the normalizer
of the respective MAD-group is isomorphic to $SL(2,\mathbb Z_n)$,
which is isomorphic to $Sp(2,\mathbb Z_n)$, for any positive
integer $n>1$. This suggests that in the case of a MAD-group
formed by inner automorphisms generated by $\underbrace{{\cal P}_n
\otimes {\cal P}_n\otimes\ldots\otimes{\cal
P}_n}_{k-\mathrm{times}}$, the normalizer may be isomorphic to
$Sp(2k,\mathbb Z_n)$.

Let us mention that the normalizer has not yet been described for
any MAD-group containing outer automorphisms.


\section*{Appendix A}
 This section contains a proof of Proposition \ref{theo_inverse
 images}. The matrices $D_j$'s considered in the proof are defined
 by (\ref{maticeD}).

\pf In order to prove that an automorphism $\varphi_j=Ad_{B_j}$ is
an inverse image of $D_j$, we must express the action of
$\varphi_j$ on the basis elements $P\otimes I, Q\otimes I,
I\otimes P, I\otimes Q$ of $\mathcal G$ again in terms of
$P\otimes I,$ $Q\otimes I,$ $I\otimes P,$ $I\otimes Q$. The
coefficients $a_{kl}$ describing the action of $\varphi_j$ (as
introduced in (\ref{mapping})) then form the matrix $D_j$. In the
following we set the four matrices $B_j$, and verify that each
satisfies the equation $D_j=C(Ad_{B_j})$.

Throughout the proof, we use the coefficient $\omega$, which is,
as defined previously, the $n$-th root of unity: $\omega=\omega_n
= e^{\frac{2\pi i}{n}}$. We also shorten the notation of $I_n$ to
$I$.

And, finally, the elements of matrices $P$, $Q$ (whose indices are
also counted modulo $n$) can be written in terms of the Kronecker
symbol as
\begin{equation}\label{PQ elements}
P_{ij}=\delta_{ij}\omega^j, \qquad Q_{ij}=\delta_{i(j-1)}, \qquad
Q^\top_{ij}=\delta_{i(j+1)}.
\end{equation}

\medskip
\noindent{\bf 1)} \ \  We define $\varphi_1=Ad_{B_1}$, where

\bigskip

\fbox{\quad\parbox{0.80\textwidth}{\vskip-0.3cm
$${B_1=\widetilde{B}_1\otimes I, \quad
\widetilde{B}_1=\mathrm{diag}(b_0, b_1, \ldots, b_{(n-1)}), \quad
b_j=\varepsilon^j\omega^{\frac{j(j-1)}{2}}, \quad
\varepsilon=\omega^{-\frac{n-1}{2}}}.$$ }\quad }


\bigskip

\noindent As $\widetilde{B}_1$, $\widetilde{B}_1^{-1}$, and $P$
are diagonal, they all mutually commute, and thus
$B^{-1}_1(P\otimes I)B_1=(\widetilde{B}_1\otimes I)^{-1}(P\otimes
I)(\widetilde{B}_1\otimes I)=(\widetilde{B}_1^{-1}\otimes
I)(P\widetilde{B}_1\otimes
I)=(\widetilde{B}_1^{-1}P\widetilde{B}_1)\otimes
I=(\widetilde{B}_1^{-1}\widetilde{B}_1P)\otimes I=P\otimes I$. In
other words, $$\varphi_1Ad_{P\otimes
I}\varphi_1^{-1}=Ad_{B_1}Ad_{P\otimes
I}Ad_{B_1^{-1}}=Ad_{B_1^{-1}(P\otimes I) B_1}=Ad_{P\otimes I},$$
which means, according to the definition of $C(\varphi_1)$, that
the first column of the matrix $C(\varphi_1)$ is\\
$\bullet$\ \
\underline{$(a_{11},a_{21},a_{31},a_{41})^\top=(1,0,0,0)^\top$}


\bigskip

\noindent Now we apply $\varphi_1$ on the second generator of the
group $\mathcal G$, which is the inner automorphism defined by the
matrix $Q\otimes I$. In fact, we need to express
$B_1^{-1}(Q\otimes I)B_1$ in terms of the basis matrices $P\otimes
I, Q\otimes I, I\otimes P, I\otimes Q$. Using the notation of
elements of $P$ and $Q$ introduced in (\ref{PQ elements}), we
obtain
\begin{eqnarray*}
(\widetilde{B}_1^{-1}Q\widetilde{B}_1)_{ij}&=&
\sum_{k=0}^{n-1}\sum_{l=0}^{n-1}(\widetilde{B}_1^{-1})_{ik}Q_{kl}(\widetilde{B}_1)_{lj}\\
&=&\sum_{k=0}^{n-1}\sum_{l=0}^{n-1}\delta_{ik}\varepsilon^{-k}\omega^{-\frac{k(k-1)}{2}}
\delta_{k(l-1)}\delta_{lj}\varepsilon^j\omega^{\frac{j(j-1)}{2}}\\
&=&\sum_{k=0}^{n-1}\delta_{ik}\varepsilon^{j-k}\omega^{\frac{j(j-1)-k(k-1)}{2}}\delta_{k(j-1)}=
\varepsilon\delta_{i(j-1)}\omega^{j-1}\\
(PQ)_{ij}&=&\sum_{k=0}^{n-1}P_{ik}Q_{kj}=\sum_{k=0}^{n-1}\delta_{ik}\omega^k\delta_{k(j-1)}=
\delta_{i(j-1)}\omega^{j-1}.
\end{eqnarray*}

\noindent We see that the matrix $B_1^{-1}(Q\otimes I)B_1$ is just
an $\varepsilon$ multiple of $PQ$, and it follows that
$$\varphi_1Ad_{Q\otimes I}\varphi_1^{-1}=Ad_{B_1}Ad_{Q\otimes
I}Ad_{B_1^{-1}}=Ad_{B_1^{-1}(Q\otimes I)
B_1}=Ad_{(\widetilde{B}_1^{-1}Q\widetilde{B}_1)\otimes I}
=Ad_{(\varepsilon PQ)\otimes I}=Ad_{PQ\otimes I}.$$

\noindent The second column of the matrix $C(\varphi_1)$ is thus
equal to
\\
$\bullet$\ \
\underline{$(a_{12},a_{22},a_{32},a_{42})^\top=(1,1,0,0)^\top$.}

\bigskip

\noindent By simple matrix multiplication, we see that
$B_1^{-1}(I\otimes P )B_1=(\widetilde{B}_1^{-1}\otimes I)(I\otimes
P)(\widetilde{B}_1\otimes
I)=(\widetilde{B}_1^{-1}\widetilde{B}_1)\otimes P=I\otimes P$;
which means $\varphi_1 Ad_{I\otimes
P}\varphi_1^{-1}=Ad_{B_1}Ad_{I\otimes
P}Ad_{B_1^{-1}}=Ad_{B_1^{-1}(I\otimes P)B_1}=Ad_{I\otimes P}$, and
therefore\\
$\bullet$\ \
\underline{$(a_{13},a_{23},a_{33},a_{43})^\top=(0,0,1,0)^\top$.}\\


\bigskip

\noindent Analogously, putting $Q$ on the place of $P$, we have
$B_1^{-1}(I\otimes Q )B_1=(\widetilde{B}_1^{-1}\otimes I)(I\otimes
Q)(\widetilde{B}_1\otimes
I)=(\widetilde{B}_1^{-1}\widetilde{B}_1)\otimes Q=I\otimes Q$;
which means $\varphi_1 Ad_{I\otimes
Q}\varphi_1^{-1}=Ad_{B_1}Ad_{I\otimes
Q}Ad_{B_1^{-1}}=Ad_{B_1^{-1}(I\otimes Q)B_1}=Ad_{I\otimes Q}$, and
therefore\\
$\bullet$\ \
\underline{$(a_{14},a_{24},a_{34},a_{44})^\top=(0,0,0,1)^\top$.}\\

Thus, we have shown that $$C(\varphi_1)=D_1 = \left(
\begin{smallmatrix}
 1& 1& 0& 0\\
 0& 1& 0& 0\\
 0& 0& 1& 0\\
 0& 0& 0& 1\\
  \end{smallmatrix}
 \right).$$


\noindent{\bf 2)} \ \  We define $\varphi_2=Ad_{B_2}$, where


\bigskip

 \fbox{\quad\parbox{0.60\textwidth}{\vskip-0.2cm
$${B_2=\widetilde{B}_2\otimes I, \quad
(\widetilde{B}_2)_{ij}=\omega^{ij}, \quad
i,j=0,1,\ldots,n-1}.$$}\quad }

\bigskip

Note that $\widetilde{B}_2$ is the famous Sylvester matrix.

\noindent In preparation for describing the action of $Ad_{B_2}$,
we apply the matrix $\widetilde{B}_2$ on $P$, $Q$, and $Q^\top$
from both right and left:
\begin{eqnarray*}
(P\widetilde{B}_2)_{ij}&=&\sum_{k=0}^{n-1}P_{ik}(\widetilde{B}_2)_{kj}=
\sum_{k=0}^{n-1}\delta_{ik}\omega^k\omega^{kj}=\omega^i\omega^{ij}=\omega^{i(j+1)}\\
(\widetilde{B}_2Q^\top)_{ij}&=&\sum_{k=0}^{n-1}(\widetilde{B}_2)_{ik}(Q^\top)_{kj}=
\sum_{k=0}^{n-1}\omega^{ik}\delta_{k(j+1)}=\omega^{i(j+1)}\\
(Q\widetilde{B}_2)_{ij}&=&\sum_{k=0}^{n-1}Q_{ik}(\widetilde{B}_2)_{kj}=
\sum_{k=0}^{n-1}\delta_{i(k-1)}\omega^{kj}=\omega^{(i+1)j}\\
(\widetilde{B}_2P)_{ij}&=&\sum_{k=0}^{n-1}(\widetilde{B}_2)_{ik}P_{kj}=
\sum_{k=0}^{n-1}\omega^{ik}\delta_{kj}\omega^j=\omega^{ij}\omega^j=\omega^{(i+1)j}.\\
\end{eqnarray*}

\noindent We easily conclude that
\begin{eqnarray*}
P\widetilde{B}_2=\widetilde{B}_2Q^\top\qquad&\Rightarrow&\qquad
\widetilde{B}_2^{-1}P\widetilde{B}_2=Q^\top=Q^{-1},\\
Q\widetilde{B}_2=\widetilde{B}_2P\qquad&\Rightarrow&\qquad
\widetilde{B}_2^{-1}Q\widetilde{B}_2=P.
\end{eqnarray*}
Using these relations, the way to find the coefficients of
$C(\varphi_2)=C(Ad_{B_2})$ is quite straightforward:

\bigskip

\noindent $B_2^{-1}(P\otimes I)B_2=(\widetilde{B}_2\otimes
I)^{-1}(P\otimes I)(\widetilde{B}_2\otimes
I)=(\widetilde{B}_2^{-1}\otimes I)(P\widetilde{B}_2\otimes
I)=(\widetilde{B}_2^{-1}P\widetilde{B}_2)\otimes I=Q^{-1}\otimes
I$. Therefore, $\varphi_2Ad_{P\otimes
I}\varphi_2^{-1}=Ad_{B_2^{-1}(P\otimes I)B_2}=Ad_{(Q\otimes
I)^{-1}}$, and we have found the coefficients\\
$\bullet$\ \
\underline{$(a_{11},a_{21},a_{31},a_{41})^\top=(0,-1,0,0)^\top$.}

\bigskip

\noindent $B_2^{-1}(Q\otimes I)B_2=(\widetilde{B}_2\otimes
I)^{-1}(Q\otimes I)(\widetilde{B}_2\otimes
I)=(\widetilde{B}_2^{-1}\otimes I)(Q\widetilde{B}_2\otimes
I)=(\widetilde{B}_2^{-1}Q\widetilde{B}_2)\otimes I=P\otimes I$.
Therefore, $\varphi_2Ad_{Q\otimes
I}\varphi_2^{-1}=Ad_{B_2^{-1}(Q\otimes I)B_2}=Ad_{(P\otimes I)}$,
and the respective coefficients of $D_2$ are\\
$\bullet$\ \
\underline{$(a_{12},a_{22},a_{32},a_{42})^\top=(1,0,0,0)^\top$.}


\bigskip

\noindent $B_2^{-1}(I\otimes P)B_2=(\widetilde{B}_2\otimes
I)^{-1}(I\otimes P)(\widetilde{B}_2\otimes
I)=(\widetilde{B}_2^{-1}\otimes I)(\widetilde{B}_2\otimes
P)=(\widetilde{B}_2^{-1}\widetilde{B}_2)\otimes P=I\otimes P$. In
this case, $\varphi_2Ad_{I\otimes
P}\varphi_2^{-1}=Ad_{B_2^{-1}(I\otimes P)B_2}=Ad_{(I\otimes P)}$,
and the respective coefficients of $D_2$ are\\
$\bullet$\ \
\underline{$(a_{13},a_{23},a_{33},a_{43})^\top=(0,0,1,0)^\top$.}

\bigskip

\noindent $B_2^{-1}(I\otimes Q)B_2=(\widetilde{B}_2\otimes
I)^{-1}(I\otimes Q)(\widetilde{B}_2\otimes
I)=(\widetilde{B}_2^{-1}\otimes I)(\widetilde{B}_2\otimes
Q)=(\widetilde{B}_2^{-1}\widetilde{B}_2)\otimes Q=I\otimes Q$. So
lastly, we obtain $\varphi_2Ad_{I\otimes
Q}\varphi_2^{-1}=Ad_{B_2^{-1}(I\otimes Q)B_2}=Ad_{(I\otimes Q)}$,
and obviously,\\
$\bullet$\ \
\underline{$(a_{14},a_{24},a_{34},a_{44})^\top=(0,0,0,1)^\top$.}

\bigskip

Thus, the matrix $C(\varphi_2)$ is equal to
$$C(\varphi_2)=D_2 = \left(
\begin{smallmatrix}
 0& 1& 0& 0\\
 -1& 0& 0& 0\\
 0& 0& 1& 0\\
 0& 0& 0& 1\\
  \end{smallmatrix}
 \right).$$

\noindent{\bf 3)} \ \  We define $\varphi_3=Ad_{B_3}$, where


\bigskip

\fbox{\quad\parbox{.85\textwidth}{\vskip-0.3cm
$${(B_3)_{pq}=\delta_{p_1q_2}\delta_{p_2q_1},\quad p=p_1
n+p_2, q=q_1 n+q_2,\quad p_1,p_2,q_1,q_2\in\{0,1,\ldots,n-1\}}.$$
}\quad }


\bigskip

\noindent (One can easily verify by a direct calculation that
$B_3^{-1}=B_3$.)

In order to satisfy the relation $C(\varphi_3)=C(Ad_{B_3})=D_3$,
it is sufficient to show that the matrix $B_3$ fulfils the
following equations:
\begin{eqnarray*}
B_3^{-1}(P\otimes I)B_3=I\otimes
P&\qquad&B_3^{-1}(I\otimes P)B_3=P\otimes I\\
B_3^{-1}(Q\otimes I)B_3=I\otimes Q&\qquad&B_3^{-1}(I\otimes
Q)B_3=Q\otimes I.
\end{eqnarray*}

In fact, we have found $B_3$ such that even a more general
relation is satisfied:
\begin{equation}\label{interchange of KL}
B_3^{-1}(K\otimes L)B_3=L\otimes K \mathrm{\ for \ any \ matrices
\ } K,L\in\mathbb C^{n\times n}.
\end{equation}
The matrix elements of $B_3$ are $\delta_{p_1q_2}\delta_{p_2q_1}$
as introduced above. In order to prove the equation
(\ref{interchange of KL}), it is sufficient to express the
$(pq)$-th element of the tensor product $K\otimes L$ as $(K\otimes
L)_{pq}=K_{p_1q_1}L_{p_2q_2}$, and proceed by
\begin{eqnarray*}
[B_3^{-1} (K\otimes L) B_3]_{pq}&=&[B_3^{-1}(K\otimes L)B_3
]_{(p_1n+p_2)(q_1n+q_2)}\\
&=&\sum_{r_1,r_2=0}^{n-1}(B_3^{-1})_{(p_1n+p_2)(r_1n+r_2)}[(K\otimes
L)B_3]_{(r_1n+r_2)(q_1n+q_2)}\\
&=&\sum_{r_1,r_2=0}^{n-1}\sum_{s_1,s_2=0}^{n-1}(B_3^{-1})_{(p_1n+p_2)(r_1n+r_2)}(K\otimes
L)_{(r_1n+r_2)(s_1n+s_2)}(B_3)_{(s_1n+s_2)(q_1n+q_2)}\\
&=&\sum_{r_1,r_2=0}^{n-1}\sum_{s_1,s_2=0}^{n-1}
\delta_{p_1r_2}\delta_{p_2r_1}K_{r_1s_1}L_{r_2s_2}\delta_{s_1q_2}\delta_{s_2q_1}\\
&=&K_{p_2q_2}L_{p_1q_1}=L_{p_1q_1}K_{p_2q_2}=(L\otimes K)_{pq}.
\end{eqnarray*}

\bigskip

\noindent{\bf 4)} \ \  We define $\varphi_4=Ad_{B_4}$, where

\bigskip

\fbox{\quad\parbox{.88\textwidth}{\vskip-0.3cm
$${(B_4)_{pq}=\delta_{(p_1-p_2)q_1}\delta_{p_2q_2},\quad
p=p_1 n+p_2, q=q_1 n+q_2,\quad
p_1,p_2,q_1,q_2\in\{0,1,\ldots,n-1\}}.$$ }\quad }


\bigskip
\noindent  Remember that, with the matrix coefficients
$p_1,p_2,q_1,q_2$, we count modulo $n$.

We first express the matrices relevant for the proof by means of
their $pq$-th elements:
\begin{eqnarray*}
(P\otimes I)_{pq}&=&\omega^{p_1}\delta_{p_1q_1}\delta_{p_2q_2}\\
(P\otimes
P)_{pq}&=&\omega^{p_1+p_2}\delta_{p_1q_1}\delta_{p_2q_2}\\
(I\otimes P)_{pq}&=&\omega^{p_2}\delta_{p_1q_1}\delta_{p_2q_2}\\
(Q\otimes I)_{pq}&=&\delta_{(p_1+1)q_1}\delta_{p_2q_2}\\
(Q^{-1}\otimes Q)_{pq}&=&\delta_{p_1(q_1+1)}\delta_{(p_2+1)q_2}\\
(I\otimes Q)_{pq}&=&\delta_{p_1q_1}\delta_{(p_2+1)q _2}.\\
\end{eqnarray*}

Then, we proceed by showing that the elements of matrix $D_4$
indeed reflect the action of the automorphism
$\varphi_4=Ad_{B_4}$:


\begin{eqnarray*}
[(P\otimes I)B_4]_{pq}&=&\sum_{r_1,r_2=0}^{n-1}(P\otimes
I)_{(p_1n+p_2)(r_1n+r_2)}(B_4)_{(r_1n+r_2)(q_1n+q_2)}\\
&=&\sum_{r_1,r_2=0}^{n-1}\omega^{p_1}
\delta_{p_1r_1}\delta_{p_2r_2}
\delta_{(r_1-r_2)q_1}\delta_{r_2q_2}\\
&=&\omega^{p_1}\delta_{(p_1-p_2)q_1}\delta_{p_2q_2},
\end{eqnarray*}
\begin{eqnarray*}
[B_4(P\otimes
P)]_{pq}&=&\sum_{r_1,r_2=0}^{n-1}(B_4)_{(p_1n+p_2)(r_1n+r_2)}(P\otimes
P)_{(r_1n+r_2)(q_1n+q_2)}\\
&=&\sum_{r_1,r_2=0}^{n-1}\delta_{(p_1-p_2)r_1}\delta_{p_2r_2}
\omega^{r_1+r_2}\delta_{r_1q_1}\delta_{r_2q_2}\\
&=&\delta_{(p_1-p_2)q_1}\delta_{p_2q_2}\omega^{q_1+q_2}=\omega^{p_1}\delta_{(p_1-p_2)q_1}\delta_{p_2q_2}.
\end{eqnarray*}
Thus we have shown that $(P\otimes I)B_4=B_4(P\otimes P)$, and
consequently, $B_4^{-1}(P\otimes I)B_4=P\otimes P$, which gives
the coefficients\\
$\bullet$\ \
\underline{$(a_{11},a_{21},a_{31},a_{41})^\top=(1,0,1,0)^\top$.}


\begin{eqnarray*}
[(Q\otimes I)B_4]_{pq}&=&\sum_{r_1,r_2=0}^{n-1}(Q\otimes
I)_{(p_1n+p_2)(r_1n+r_2)}(B_4)_{(r_1n+r_2)(q_1n+q_2)}\\
&=&\sum_{r_1,r_2=0}^{n-1}\delta_{(p_1+1)r_1}\delta_{p_2r_2}
\delta_{(r_1-r_2)q_1}\delta_{r_2q_2}\\
&=&\delta_{p_2q_2}\delta_{(p_1+1)(q_1+q_2)},
\end{eqnarray*}
\begin{eqnarray*}
[B_4(Q\otimes
I)]_{pq}&=&\sum_{r_1,r_2=0}^{n-1}(B_4)_{(p_1n+p_2)(r_1n+r_2)}(Q\otimes
I)_{(r_1n+r_2)(q_1n+q_2)}\\
&=&\sum_{r_1,r_2=0}^{n-1}\delta_{(p_1-p_2)r_1}\delta_{p_2r_2}
\delta_{(r_1+1)q_1}\delta_{r_2q_2}\\
&=&\delta_{p_2q_2}\delta_{(p_1-p_2)(q_1-1)}=\delta_{p_2q_2}\delta_{(p_1+1)(q_1+q_2)}.
\end{eqnarray*}
Combining the two equations, we obtain $(Q\otimes
I)B_4=B_4(Q\otimes I)$, and $B_4^{-1}(Q\otimes I)B_4$. In terms of
the matrix coefficients of $D_4$, it writes as\\
$\bullet$\ \
\underline{$(a_{12},a_{22},a_{32},a_{42})^\top=(0,1,0,0)^\top$}.


\begin{eqnarray*}
[(I\otimes P)B_4]_{pq}&=&\sum_{r_1,r_2=0}^{n-1}(I\otimes
P)_{(p_1n+p_2)(r_1n+r_2)}(B_4)_{(r_1n+r_2)(q_1n+q_2)}\\
&=&\sum_{r_1,r_2=0}^{n-1}\omega^{p_2}
\delta_{p_1r_1}\delta_{p_2r_2}
\delta_{(r_1-r_2)q_1}\delta_{r_2q_2}\\
&=&\omega^{p_2}\delta_{(p_1-p_2)q_1}\delta_{p_2q_2},
\end{eqnarray*}
\begin{eqnarray*}
[B_4(I\otimes
P)]_{pq}&=&\sum_{r_1,r_2=0}^{n-1}(B_4)_{(p_1n+p_2)(r_1n+r_2)}(I\otimes
P)_{(r_1n+r_2)(q_1n+q_2)}\\
&=&\sum_{r_1,r_2=0}^{n-1}\delta_{(p_1-p_2)r_1}\delta_{p_2r_2}
\omega^{r_2}\delta_{r_1q_1}\delta_{r_2q_2}\\
&=&\omega^{p_2}\delta_{(p_1-p_2)q_1}\delta_{p_2q_2}.
\end{eqnarray*}
>From $(I\otimes P)B_4=B_4(I\otimes P)$ we obtain
$B_4^{-1}(I\otimes P)B_4=I\otimes P$, i.e. the matrix $I\otimes P$
remains intact by the action of $Ad_{B_4}$, which means that\\
$\bullet$\ \
\underline{$(a_{13},a_{23},a_{33},a_{43})^\top=(0,0,1,0)^\top$}.


\begin{eqnarray*}
[(I\otimes Q)B_4]_{pq}&=&\sum_{r_1,r_2=0}^{n-1}(I\otimes
Q)_{(p_1n+p_2)(r_1n+r_2)}(B_4)_{(r_1n+r_2)(q_1n+q_2)}\\
&=&\sum_{r_1,r_2=0}^{n-1}\delta_{p_1r_1}\delta_{(p_2+1)r_2}
\delta_{(r_1-r_2)q_1}\delta_{r_2q_2}\\
&=&\delta_{(p_2+1)q_2}\delta_{p_1(q_1+q_2)},
\end{eqnarray*}
\begin{eqnarray*}
[B_4(Q^{-1}\otimes
Q)]_{pq}&=&\sum_{r_1,r_2=0}^{n-1}(B_4)_{(p_1n+p_2)(r_1n+r_2)}(Q^{-1}\otimes
Q)_{(r_1n+r_2)(q_1n+q_2)}\\
&=&\sum_{r_1,r_2=0}^{n-1}\delta_{(p_1-p_2)r_1}\delta_{p_2r_2}
\delta_{r_1(q_1+1)}\delta_{(r_2+1)q_2}\\
&=&\delta_{(p_2+1)q_2}\delta_{(p_1-p_2)(q_1+1)}=\delta_{(p_2+1)q_2}\delta_{p_1(q_1+q_2)}.
\end{eqnarray*}
Here, we see the action of $Ad_{B_4}$ on the last of the four
matrices generating $\mathcal G$: $(I\otimes
Q)B_4=B_4(Q^{-1}\otimes Q)$ is equivalent to $B_4^{-1}(I\otimes
Q)B_4=Q^{-1}\otimes Q$. As a result,\\
$\bullet$\ \
\underline{$(a_{14},a_{24},a_{34},a_{44})^\top=(0,-1,0,1)^\top$}.

\bigskip

Thus, we have shown that $$C(\varphi_4)=D_4 = \left(
\begin{smallmatrix}
 1& 0& 0& 0\\
 0& 1& 0& -1\\
 1& 0& 1& 0\\
 0& 0& 0& 1\\
  \end{smallmatrix}
 \right).$$

\pfk

\section*{Appendix B}

\begin{thm} Let $n$ be a prime. Then the four matrices
$$ D_1 = \left( \begin{smallmatrix}
 1& 1& 0& 0\\
 0& 1& 0& 0\\
 0& 0& 1& 0\\
 0& 0& 0& 1\\
  \end{smallmatrix}
 \right), \ \
D_2 = \left( \begin{smallmatrix}
 0&1& 0& 0\\
 -1& 0& 0& 0\\
 0& 0& 1& 0\\
 0& 0& 0& 1\\
  \end{smallmatrix}
 \right), \ \
 D_3 = \left( \begin{smallmatrix}
 0& 0& 1& 0\\
 0& 0& 0& 1\\
 1& 0& 0& 0\\
 0& 1& 0& 0\\
  \end{smallmatrix}
 \right), \ \  {\rm and} \ \
 D_4 = \left( \begin{smallmatrix}
 1& 0& 0& 0\\
 0& 1& 0& -1\\
 1& 0& 1& 0\\
 0& 0& 0& 1\\
  \end{smallmatrix}
 \right) $$
generate the group  $Sp(4, \mathbb{F}_n)$. \label{theo_D1234}
\end{thm}
For a better overview, we first sketch the main framework of the
proof, and only afterwards we prove the individual steps in
detail.

\begin{pozn}
The fact that $n$ is a prime ensures that $\mathbb Z_n$ is a field
$\mathbb F_n$, and consequently that the set $Sp(4,\mathbb
F_n)=\{X\in \mathbb{Z}_n^{4\times 4}\,|\,X^\top\! JX =
 J\},$ where $J= \left(\begin{smallmatrix}
      0&1\\-1&0\end{smallmatrix}\right) \oplus \left(\begin{smallmatrix}
      0&1\\-1&0\end{smallmatrix}\right)$,
is a group.
\end{pozn}

\begin{pozn}\label{sl2}  It is a well known fact that, for any integer $n$, the
two matrices $\left(\begin{smallmatrix} 1&1\\0&1
\end{smallmatrix}\right)$ and $\left(\begin{smallmatrix} 0&1\\-1&0
\end{smallmatrix}\right)$ generate the group $SL(2, \mathbb{Z}_n)
= \{ A\in \mathbb{Z}_n^{2\times 2}\ |\  \det A=1\} $ (see
\cite{HaPaPeTo}).
\end{pozn}

\begin{pozn} It is easy to verify that $D_1, D_2, D_3$, and $D_4$
belong to $Sp(4,\mathbb{F}_n)$.
\end{pozn}

For a group ${\cal K}$ and any group elements $k_1,k_2,\ldots k_r
\in \cal K$, we denote by $\langle k_1,\ldots , k_r\rangle$ the
smallest subgroup of $\cal K$ containing $k_1, \ldots ,k_r$.

Under the framework of this notation and Remark \ref{sl2}, we have
$$\langle D_1,D_2\rangle = \Bigl\{\Bigl(\begin{array}{cc}
A&0\\
0&I_2
\end{array}\Bigr)\Bigl| \ A\in Sl(2, \mathbb{Z}_n)\Bigr\},$$
$$\langle D_1,D_2, D_3\rangle = \Bigl\{\Bigl(\begin{array}{cc}
A&0\\
0&B
\end{array}\Bigr), \Bigl(\begin{array}{cc}
0&A\\
B&0
\end{array}\Bigr)\Bigl| \ A,B\in Sl(2, \mathbb{Z}_n)\Bigr\}.$$

Our aim is to prove that $\langle D_1,D_2, D_3, D_4\rangle =
Sp(4,\mathbb{Z}_n)$. For this purpose, it is enough to verify
statements of the next two steps:

\begin{enumerate}
    \item[{\it Step 1.}] Let us note $\mathcal{H}:=\langle
    D_1,D_2,D_3\rangle$ and $M\in Sp(4,\mathbb{F}_n),\ M\notin
    \mathcal{H}$. There exist $k\in \Z_n$ and matrices $G_1,G_2\in
    \mathcal{H}$ such that $G_1MG_2=S(k)$, where
    $$
    S(k):=\left(%
    \begin{array}{cccc}
  1 & 0 & 1 & 0 \\
  0 & k & 0 & 1-k \\
  k-1 & 0 & k & 0 \\
  0 & -1 & 0 & 1 \\
    \end{array}%
    \right).
    $$
    \item[{\it Step 2.}] The matrix $S(k)$ belongs to $\langle
    D_1,D_2,D_3,D_4\rangle$ for any $k\in \Z_n$.
\end{enumerate}

\begin{lem} Let $n$ be a prime and $A=\left(\begin{smallmatrix}
a & b \\
c & d \\
\end{smallmatrix}\right)\in \Z_n^{2\times 2}$, $A\neq \left(\begin{smallmatrix}
  0 & 0 \\
  0 & 0 \\
\end{smallmatrix} \right)$. Then there exist
\begin{enumerate}
    \item[\emph{(i)}] matrices $B,C\in SL(2,\Z_n)$ such that $BAC=\left(\begin{smallmatrix}
  1 & 0 \\
  0 & k \\
\end{smallmatrix} \right)$, where $k=\det A$;
    \item[\emph{(ii)}] matrices $D,E\in SL(2,\Z_n)$ such that $DAE=\left(\begin{smallmatrix}
  k & 0 \\
  0 & 1 \\
\end{smallmatrix} \right)$, where $k=\det A$.
\end{enumerate}
Moreover, if $\det A=k\neq 0$, then we have $C=E=I_2$.
\label{lemmaBCDE}
\end{lem}

\pf a) Firstly, we consider the case $\det A=k \neq 0$.
\begin{enumerate}
    \item[(i)] We assume, without loss of generality, that $a\neq
    0$ (otherwise, we would consider matrix $A\left(\begin{smallmatrix}
  0 & 1 \\
  -1 & 0 \\
\end{smallmatrix} \right)= \left(\begin{smallmatrix}
  -b & a \\
  -d & c \\
\end{smallmatrix} \right)= \left(\begin{smallmatrix}
  -b & 0 \\
  -d & c \\
\end{smallmatrix} \right)$, as $b\neq 0$ when $a=0$). The desired
matrices $B$ and $C$ are as follows:
$$
\begin{array}{rcl}
B&=&
B_3B_2B_1=\left(%
\begin{array}{cc}
  1 & -ba^{-1}k^{-1} \\
  0 & 1 \\
\end{array}%
\right)\left(%
\begin{array}{cc}
  1 & 0 \\
  -ac & 1 \\
\end{array}%
\right)\left(%
\begin{array}{cc}
  a^{-1} & 0 \\
  0 & a \\
\end{array}%
\right),\\ C&=&I_2, \\*[2ex] BAC&=&
B_3B_2\left(%
\begin{array}{cc}
  a^{-1} & 0 \\
  0 & a \\
\end{array}%
\right)\left(%
\begin{array}{cc}
  a & b \\
  c & d \\
\end{array}%
\right)=B_3 \left(%
\begin{array}{cc}
  1 & 0 \\
  -ac & 1 \\
\end{array}%
\right)\left(%
\begin{array}{cc}
  1 & a^{-1}b \\
  ac & ad \\
\end{array}%
\right)\\*[2ex] &=&\left(%
\begin{array}{cc}
  1 & -ba^{-1}k^{-1} \\
  0 & 1 \\
\end{array}%
\right)\left(%
\begin{array}{cc}
  1 & a^{-1}b \\
  0 & k \\
\end{array}%
\right)=\left(%
\begin{array}{cc}
  1 & 0 \\
  0 & k \\
\end{array}%
\right).
\end{array}
$$
\item[(ii)] Alternatively, we multiply the result by one more
matrix, in order to obtain the desired matrix
$\left(\begin{smallmatrix}
  k & 0 \\
  0 & 1 \\
\end{smallmatrix} \right)$:
$$
\begin{array}{l}
D=\left(%
\begin{array}{cc}
  k & 0 \\
  0 & k^{-1} \\
\end{array}%
\right)B,\\ E=C=I_2,\\*[2ex] DAE=\left(%
\begin{array}{cc}
  k & 0 \\
  0 & k^{-1} \\
\end{array}%
\right)BAC=\left(%
\begin{array}{cc}
  k & 0 \\
  0 & k^{-1} \\
\end{array}%
\right)\left(%
\begin{array}{cc}
  1 & 0 \\
  0 & k \\
\end{array}%
\right)=\left(%
\begin{array}{cc}
  k & 0 \\
  0 & 1 \\
\end{array}%
\right).
\end{array}
$$
\end{enumerate}
\noindent b) Secondly, we have the situation $\det A=k=0$, but
still $A\neq 0$.

\begin{enumerate}
\item[(i)] If $a\neq 0$, we put
$$
\begin{array}{rcl}
B&=&B_2B_1,\\ C&=&\left(%
\begin{array}{cc}
  1 & -a^{-1}b \\
  0 & 1 \\
\end{array}%
\right) ,\\*[2ex]BAC&=&B_2\left(%
\begin{array}{cc}
  a^{-1} & 0 \\
  0 & a \\
\end{array}%
\right)\left(%
\begin{array}{cc}
  a & b \\
  c & d \\
\end{array}%
\right)C=
\left(%
\begin{array}{cc}
  1 & 0 \\
  -ac & 1 \\
\end{array}%
\right)\left(%
\begin{array}{cc}
  1 & a^{-1}b \\
  ac & ad \\
\end{array}%
\right)C\\*[2ex] &=&\left(%
\begin{array}{cc}
  1 & a^{-1}b \\
  0 & k \\
\end{array}%
\right)\left(%
\begin{array}{cc}
  1 & -a^{-1}b \\
  0 & 1 \\
\end{array}%
\right)=\left(%
\begin{array}{cc}
  1 & 0 \\
  0 & k \\
\end{array}%
\right)=\left(%
\begin{array}{cc}
  1 & 0 \\
  0 & 0 \\
\end{array}%
\right).
\end{array}
$$

If $a=0$, then at least one of the remaining three matrix elements
is non-zero. Thus, we can analogously work with one of the
matrices
\begin{equation}
A\left(%
\begin{array}{cc}
  0 & 1 \\
  -1 & 0 \\
\end{array}%
\right),\ \left(%
\begin{array}{cc}
  0 & 1 \\
  -1 & 0 \\
\end{array}%
\right)A,\ \left(%
\begin{array}{cc}
  0 & 1 \\
  -1 & 0 \\
\end{array}%
\right)A\left(%
\begin{array}{cc}
  0 & 1 \\
  -1 & 0 \\
\end{array}
\right), \label{eq_3matrices}
\end{equation} namely with the one whose element in the
first row and first column is non-zero.

\item[(ii)] Again, without loss of generality, we assume that
$d\neq 0$ (otherwise we would transform $A$ into one of the three
matrices given in (\ref{eq_3matrices}); one of which would have a
non-zero element in its second row and second column). We put

$$
\begin{array}{rcl}
D&=&D_2D_1=\left(%
\begin{array}{cc}
  1 & -bd \\
  0 & 1 \\
\end{array}%
\right)\left(%
\begin{array}{cc}
  d & 0 \\
  0 & d^{-1} \\
\end{array}%
\right),\\

E&=&\left(%
\begin{array}{cc}
  1 & 0 \\
  -cd^{-1} & 1 \\
\end{array}%
\right),\\*[2ex] DAE&=&D_2\left(%
\begin{array}{cc}
  d & 0 \\
  0 & d^{-1} \\
\end{array}%
\right)\left(%
\begin{array}{cc}
  a & b \\
  c & d \\
\end{array}%
\right)E=
\left(%
\begin{array}{cc}
  1 & -bd \\
  0 & 1 \\
\end{array}%
\right)\left(%
\begin{array}{cc}
  da & db \\
  d^{-1}c & 1 \\
\end{array}%
\right)E \\*[2ex] &=& \left(%
\begin{array}{cc}
  k & 0 \\
  d^{-1}c & 1 \\
\end{array}%
\right)\left(%
\begin{array}{cc}
  1 & 0 \\
  -cd^{-1} & 1 \\
\end{array}%
\right)=
\left(%
\begin{array}{cc}
  k & 0 \\
  0 & 1 \\
\end{array}%
\right)=\left(%
\begin{array}{cc}
  0 & 0 \\
  0 & 1 \\
\end{array}%
\right).
\end{array}
$$
\end{enumerate}
\pfk

\noindent \emph{Proof of Step 1}. Let us express a matrix $M\in
Sp(4,\mathbb{F}_n)$ in blocks:
$$
M=\left(%
\begin{array}{cc}
  M_{11} & M_{12} \\
  M_{21} & M_{22} \\
\end{array}%
\right), \hbox{where } M_{ij} \in \Z_n^{2\times 2}.$$ The equality
(\ref{(II)}) means that $\det M_{12}+\det M_{22}=1$. We denote
$\det M_{22}=k$, and consequently $\det M_{12}=1-k$.
\begin{enumerate}
    \item[(i)] Let us assume that $k\neq 0$. We take matrices
    $B,C\in SL(2,\Z_n)$ as described in Lemma \ref{lemmaBCDE},
    so as to obtain $BM_{12}C=\left(\begin{smallmatrix}
  1 & 0 \\
  0 & 1-k \\
\end{smallmatrix} \right)$. Afterwards, we apply Lemma
\ref{lemmaBCDE} on matrix $M_{22}C$, whose determinant is $\det
(M_{22}C)=(\det M_{22})(\det C)=k\cdot 1=k\neq 0.$ Thus, we find
$D\in SL(2,\Z_n)$ such that $DM_{22}C=\left(\begin{smallmatrix}
  k & 0 \\
  0 & 1 \\
\end{smallmatrix} \right).$ At this moment, we are able to transform $M$ by means of
matrices $F_1=\left(\begin{smallmatrix}
  B & 0 \\
  0 & D \\
\end{smallmatrix}
\right),\ F_2=\left(\begin{smallmatrix}
  I_2 & 0 \\
  0 & C \\
\end{smallmatrix} \right)$ into the following:
$$
\begin{array}{rcl}
F_1MF_2&=&\left(%
\begin{array}{cc}
  B & 0 \\
  0 & D \\
\end{array}%
\right)\left(%
\begin{array}{cc}
  M_{11} & M_{12} \\
  M_{21} & M_{22} \\
\end{array}%
\right)\left(%
\begin{array}{cc}
  I_2 & 0 \\
  0 & C \\
\end{array}%
\right)=\left(%
\begin{array}{cc}
  BM_{11} & BM_{12} \\
  DM_{21} & DM_{22} \\
\end{array}%
\right)\left(%
\begin{array}{cccc}
  I_2 & 0 \\
  0 & C \\
\end{array}%
\right) \\*[2ex] &=&\left(%
\begin{array}{cc}
  BM_{11} & BM_{12}C \\
  DM_{21} & DM_{22}C \\
\end{array}%
\right)=
\left(%
\begin{array}{cccc}
  \widetilde{m}_{11} & \widetilde{m}_{12} & 1 & 0 \\
  \widetilde{m}_{21} & \widetilde{m}_{22} & 0 & 1-k \\
  \widetilde{m}_{31} & \widetilde{m}_{32} & k & 0 \\
  \widetilde{m}_{41} & \widetilde{m}_{42} & 0 & 1 \\
\end{array}%
\right)=\widetilde{M},
\end{array}
$$ where we denote matrices $BM_{11}=\left(\begin{smallmatrix}
  \widetilde{m}_{11} & \widetilde{m}_{12} \\
  \widetilde{m}_{21} & \widetilde{m}_{22} \\
\end{smallmatrix}
\right)$ and $DM_{21}=\left(\begin{smallmatrix}
  \widetilde{m}_{31} & \widetilde{m}_{32} \\
  \widetilde{m}_{41} & \widetilde{m}_{42} \\
\end{smallmatrix} \right)$ by $\widetilde{M}_{11}$, $\widetilde{M}_{21}$, respectively.

As the matrix $\widetilde{M}$ is a multiple of $F_1, F_2,$ and
$M$, which all belong to $Sp(4,\mathbb F_n)$, then also
$\widetilde{M}$ is an element of $Sp(4,\mathbb F_n)$, and so its
elements fulfill equations (\ref{(III)})--(\ref{(VI)}):
\begin{eqnarray}
0\ =&-\widetilde{m}_{21}-k\widetilde{m}_{41}&\ ({\rm{mod}}\ n)
\nonumber\\ 0\ =&(1-k)\widetilde{m}_{11}+\widetilde{m}_{31}&\
({\rm{mod}}\ n) \nonumber\\ 0\
=&-\widetilde{m}_{22}-k\widetilde{m}_{42}&\ ({\rm{mod}}\ n)
\nonumber\\ 0\ =&(1-k)\widetilde{m}_{12}+\widetilde{m}_{32}&\
({\rm{mod}}\ n). \nonumber
\end{eqnarray}
These relations ensure that we can transform $\widetilde{M}$ into
$S(k)$ by means of matrix $N=\left(\begin{smallmatrix}
  \widetilde{m}_{11} & \widetilde{m}_{12} \\
  -\widetilde{m}_{41} & -\widetilde{m}_{42} \\
\end{smallmatrix} \right)$.
$$
\begin{array}{rcl}
\left(%
\begin{array}{cc}
  1 & 0 \\
  0 & k \\
\end{array}%
\right)N&=&\left(%
\begin{array}{cc}
  \widetilde{m}_{11} & \widetilde{m}_{12} \\
  -k\widetilde{m}_{41} & -k\widetilde{m}_{42} \\
\end{array}%
\right)=\left(%
\begin{array}{cc}
  \widetilde{m}_{11} & \widetilde{m}_{12} \\
  \widetilde{m}_{21} & \widetilde{m}_{22} \\
\end{array}%
\right)=\widetilde{M}_{11}, \\*[2ex]
\left(%
\begin{array}{cc}
  k-1 & 0 \\
  0 & -1 \\
\end{array}%
\right)N&=&\left(%
\begin{array}{cc}
  (k-1)\widetilde{m}_{11} & (k-1)\widetilde{m}_{12} \\
  \widetilde{m}_{41} & \widetilde{m}_{42} \\
\end{array}%
\right)=\left(%
\begin{array}{cc}
  \widetilde{m}_{31} & \widetilde{m}_{32} \\
  \widetilde{m}_{41} & \widetilde{m}_{42} \\
\end{array}%
\right)=\widetilde{M}_{21}.
\end{array}
$$ Therefore, $\widetilde{M}_{11}N^{-1}=\left(\begin{smallmatrix}
  1 & 0 \\
  0 & k \\
\end{smallmatrix}
\right)$, $\widetilde{M}_{21}N^{-1}=\left(\begin{smallmatrix}
  k-1 & 0 \\
  0 & -1 \\
\end{smallmatrix} \right)$, and consequently
$$
\widetilde{M}\left(%
\begin{array}{cc}
  N^{-1} & 0 \\
  0 & I_2 \\
\end{array}%
\right)=\left(%
\begin{array}{cccc}
  \widetilde{m}_{11} & \widetilde{m}_{12} & 1 & 0 \\
  \widetilde{m}_{21} & \widetilde{m}_{22} & 0 & 1-k \\
  \widetilde{m}_{31} & \widetilde{m}_{32} & k & 0 \\
  \widetilde{m}_{41} & \widetilde{m}_{42} & 0 & 1 \\
\end{array}%
\right)\left(%
\begin{array}{cc}
  N^{-1} & 0 \\
  0 & I_2 \\
\end{array}%
\right)=S(k).$$

As an element of $Sp(4,\mathbb F_n)$, the matrix $\widetilde{M}$
also fulfills equation (\ref{(I)}), which implies that $\det
\widetilde{M}_{11}+\det \widetilde{M}_{21}=1=k\det N+(1-k)\det
N=\det N$, and thus $N\in SL(2,\mathbb Z_n)$. Hence, we have found
the desired matrices $G_1=F_1=B\oplus D$, $G_2=F_2(N^{-1}\oplus
I_2)=N^{-1}\oplus C$, transforming $M$ into $G_1MG_2=S(k)$.

\item[(ii)] The case when $k=0$, we treat as follows. Firstly, we
find $D$, $E\in SL(2,\mathbb Z_n)$ such that
$DM_{22}E=\left(\begin{smallmatrix}
  k & 0 \\
  0 & 1 \\
\end{smallmatrix}
\right)$ and $B\in SL(2,\mathbb Z_n)$ such that
$BM_{21}E=\left(\begin{smallmatrix}
  1 & 0 \\
  0 & 1-k \\
\end{smallmatrix} \right)$, according to Lemma \ref{lemmaBCDE}.
Analogously to the case $k\neq 0$, we set $G_1=F_1=B\oplus D,\
G_2=F_2(N^{-1}\otimes I_2)=N^{-1}\oplus E$, and we reach the
result $G_1MG_2=S(k)$.
\end{enumerate}

\noindent \emph{Proof of Step 2}. We need to express each matrix
$S(k)$ as an element of $\langle D_1,D_2,D_3,D_4\rangle$. We begin
by showing (by induction) that
$$ D_4^j=\left(%
\begin{array}{cccc}
  1 & 0 & 0 & 0 \\
  0 & 1 & 0 & -j \\
  j & 0 & 1 & 0 \\
  0 & 0 & 0 & 1 \\
\end{array}%
\right)=
\left(%
\begin{array}{cccc}
  1 & 0 & 0 & 0 \\
  0 & 1 & 0 & -1 \\
  1 & 0 & 1 & 0 \\
  0 & 0 & 0 & 1 \\
\end{array}%
\right)\left(%
\begin{array}{cccc}
  1 & 0 & 0 & 0 \\
  0 & 1 & 0 & -j+1 \\
  j-1 & 0 & 1 & 0 \\
  0 & 0 & 0 & 1 \\
\end{array}%
\right)=D_4D_4^{j-1}. $$ We make use of the fact that
$J=\left(\begin{smallmatrix}
  0 & 1 \\
  -1 & 0 \\
\end{smallmatrix}
\right)\oplus \left(\begin{smallmatrix}
  0 & 1 \\
  -1 & 0 \\
\end{smallmatrix} \right) \in \mathcal{H}$, $J^\top \in
\mathcal{H}$, and $D_4^\top=D_3D_4D_3\in Sp(4, \mathbb F_n)$ (this
can be verified by a simple matrix multiplication ); and we
generate $S(k)$ from $D_4$ and elements from $\mathcal{H}$: $$
\begin{array}{l}
J^\top (D_4^{1-k})^\top JD_4^\top \\*[2ex]
=\left(%
\begin{array}{cccc}
  0 & -1 & 0 & 0 \\
  1 & 0 & 0 & 0 \\
  0 & 0 & 0 & -1 \\
  0 & 0 & 1 & 0 \\
\end{array}%
\right)\left(%
\begin{array}{cccc}
  1 & 0 & 1-k & 0 \\
  0 & 1 & 0 & 0 \\
  0 & 0 & 1 & 0 \\
  0 & k-1 & 0 & 1 \\
\end{array}%
\right)\left(%
\begin{array}{cccc}
  0 & 1 & 0 & 0 \\
  -1 & 0 & 0 & 0 \\
  0 & 0 & 0 & 1 \\
  0 & 0 & -1 & 0 \\
\end{array}%
\right)\left(%
\begin{array}{cccc}
  1 & 0 & 1 & 0 \\
  0 & 1 & 0 & 0 \\
  0 & 0 & 1 & 0 \\
  0 & -1 & 0 & 1 \\
\end{array}%
\right)\\*[2ex]
=\left(%
\begin{array}{cccc}
  0 & -1 & 0 & 0 \\
  1 & 0 & 1-k & 0 \\
  0 & 1-k & 0 & -1 \\
  0 & 0 & 1 & 0 \\
\end{array}%
\right)\left(%
\begin{array}{cccc}
  0 & 1 & 0 & 0 \\
  -1 & 0 & -1 & 0 \\
  0 & -1 & 0 & 1 \\
  0 & 0 & -1 & 0 \\
\end{array}%
\right)=\left(%
\begin{array}{cccc}
  1 & 0 & 1 & 0 \\
  0 & k & 0 & 1-k \\
  k-1 & 0 & k & 0 \\
  0 & -1 & 0 & 1 \\
\end{array}%
\right)=S(k).
\end{array}
$$


\section*{Acknowledgements}

Authors are indebted to the referee for his or her comments and
advices concerning relevant literature.   E.P. acknowledges
partial support by Czech Science Foundation GA \v{C}R 201/05/0169.
The research of S.T. was supported by a FQRNT Postdoctoral
Scholarship du Qu\'ebec.

\end{document}